\documentstyle[preprint,aps,eqsecnum,epsf,axodraw]{revtex}

\def\Mstar{M}
\def\Mmess{M}

\def\L{{\cal L}}

\def\O{{\cal O}}

\def\P{{\cal P}}
\def\qslash{\not{\hbox{\kern-2pt $q$}}}
\def\delslash{\not{\hbox{\kern-2pt $\partial$}}}
\def\tr{{\mbox{\rm tr}}}
\def\del{\partial}
\def\beq{\begin{equation}}
\def\eeq{\end{equation}}
\def\eeq{\end{equation}}
\def\bea{\begin{eqnarray}}
\def\eea{\end{eqnarray}}
\def\bq{\begin{quote}}
\def\eq{\end{quote}}
\def\lesssim{\mathrel{\mathpalette\vereq<}}
\def\gtrsim{\mathrel{\mathpalette\vereq>}}
\def\lsim{\mathrel{\lesssim}}
\def\gsim{\mathrel{\gtrsim}}

\makeatletter
\def\vereq#1#2{\lower3pt\vbox{\baselineskip1.5pt \lineskip1.5pt
\ialign{$\m@th#1\hfill##\hfil$\crcr#2\crcr\sim\crcr}}}
\makeatother

\begin{document}
\preprint{\vbox{\hbox{FERMILAB-Pub-99/327-T, SLAC-PUB-8291}
		\hbox{ANL-HEP-PR-110, EFI 99-48}}}

\title{Supersymmetry Breaking through Transparent Extra Dimensions}

\author{D. Elazzar Kaplan\,$^a$\thanks{\tt dkaplan@theory.uchicago.edu}, 
        Graham D. Kribs\,$^b$\thanks{\tt kribs@cmu.edu} and 
        Martin Schmaltz\,$^c$\thanks{\tt schmaltz@slac.stanford.edu}}
\address{ \vbox{\vskip 0.truecm}
  $^a$Enrico Fermi Institute, Department of Physics, \\
     University of Chicago, 5640 Ellis Avenue, Chicago, IL 60637 \\
    and Argonne National Laboratory, Argonne, IL 60439 \\
\vbox{\vskip 0.truecm}
  $^b$Department of Physics, Carnegie Mellon University, Pittsburgh, 
      PA 15213-3890 \\
\vbox{\vskip 0.truecm}
  $^c$Stanford Linear Accelerator Center \\
  Stanford University, Stanford, CA 94309}
\maketitle

\begin{abstract}%

We propose a new framework for mediating supersymmetry breaking
through an extra dimension. It predicts positive scalar masses
and solves the supersymmetric flavor problem. Supersymmetry breaks
on a ``source'' brane that is spatially separated from a parallel
brane on which the standard model matter fields and their superpartners
live. The gauge and gaugino fields propagate in the bulk, the latter
receiving a supersymmetry breaking mass from direct couplings to
the source brane. Scalar masses are suppressed at the high scale
but are generated via the renormalization group. We briefly discuss
the spectrum and collider signals for a range of compactification scales.

\end{abstract}

\section{Introduction}

Electroweak precision data indicate that the mechanism of electroweak
symmetry breaking involves a weakly coupled Higgs field. Through radiative
corrections the Higgs mass is quadratically sensitive to any scale of new
physics. It is therefore hard to understand why the Higgs mass is so much
lower than other mass scales which we believe exist in nature,
for example the Planck scale.

Low energy supersymmetry is arguably the most compelling framework for
addressing this problem: in the minimal supersymmetric standard model
(MSSM) one simply introduces superpartners which cancel the
divergences order by order in perturbation theory.
Unfortunately this solution to the hierarchy problem
introduces new problems. Accidental flavor symmetries
which suppress flavor changing neutral currents (FCNC) in the
standard model (SM) are badly broken by the supersymmetry breaking
scalar masses and $A$-terms in a generic version of the MSSM \cite{fcnc}.
Experimental limits on FCNCs force us to consider only very
special regions in parameter space where the squark and slepton mass matrices
are nearly degenerate \cite{sugra} or aligned with quark and 
lepton masses \cite{align}. Two recent proposals for the communication
of supersymmetry breaking which do give such degenerate squark
and slepton masses are gauge mediation \cite{DNS,GRreport} and 
anomaly mediation \cite{RSanom,GLMR}.\footnote{It is also possible to 
decouple the problematic flavor violating effects by by making
the first two generations of scalars heavy \cite{effsusy,negmass}.
However in practice realistic models do require some degree of
degeneracy \cite{BDetc} or alignment \cite{KKLMNR}.}

In this article we propose a new mechanism for communicating
supersymmetry breaking that leads to a distinctive spectrum of
superpartner masses. It is phenomenologically viable and respects
the approximate flavor symmetries of the SM\@. In our scenario, 
the matter fields of the MSSM (quarks, leptons,
Higgs fields and superpartners) are localized on a $3+1$
dimensional brane (the ``matter'' brane) embedded in extra dimensions.
The $SU(3)\times SU(2) \times U(1)$ gauge fields and gauginos
live in the bulk of the extra dimensions \cite{gaugeinbulk}.
Supersymmetry is broken
(dynamically) on a parallel ``source'' brane that is separated
from the matter brane in the extra dimensions \cite{CLP}. Note that in
contrast to hidden sector models, our source brane is not hidden
at all; the SM gauge fields couple directly to both branes.
This set-up leads to the following spectrum of superpartner
masses at the compactification scale: gauginos obtain masses through
their direct couplings to the supersymmetry breaking source and 
all other supersymmetry breaking masses are suppressed by the
spatial separation of the source and matter branes and/or by loop factors.
Thus after
integrating out the extra dimensional dynamics at the compactification
scale $L^{-1}$ we obtain the MSSM with the only non-negligible supersymmetry
breaking being the gaugino masses. This implies that our scenario is
very predictive since all supersymmetry breaking parameters
can be traced to a single source.

It is easy to understand that this high scale boundary condition
is also very attractive phenomenologically. The absence of
soft scalar masses and trilinear $A$ terms implies that the only
source of flavor violation is the Yukawa matrices.
This solves the supersymmetric flavor problem by a
super-GIM mechanism. Furthermore, gaugino masses contribute
to the renormalization of the scalar masses with
the correct sign to give only positive scalar squared masses.
There is one subtlety in this argument which leads to successful
radiative electroweak symmetry breaking.  Because of their strong
couplings to gluinos, the masses of colored 
scalars become large much faster than the supersymmetry breaking 
Higgs masses.  As a consequence the heavy stops running in loops
involving the large top Yukawa coupling eventually drive the up-type
Higgs (mass)$^2$ negative. Thus radiative electroweak symmetry
breaking \cite{IR} is also automatic in our framework.

For $L^{-1}$ near the Planck scale, the phenomenology of this model is
similar to that of ``no-scale'' supergravity \cite{noscale} with unified
gaugino masses.  However, in our scenario the compactification scale
is a free parameter, so the superpartner spectrum and the
associated phenomenology varies with this parameter.

In the next section we present our theoretical framework and 
discuss the coupling of bulk gauge fields to the two branes.
In Section~\ref{loops} we calculate the effective gaugino
masses and scalar masses resulting from integrating out the higher
dimensional physics for general supersymmetry breaking sectors.
As an example we then present a specific model of supersymmetry breaking.
In Section~\ref{pheno} a renormalization group
analysis is performed, determining the spectrum of superpartner masses
at the weak scale.  We find that the NLSP is nearly always the stau, 
and we show that current LEP bounds on charged sparticle masses already 
restrict a significant portion of parameter space.  Finally, the collider 
signals are briefly mentioned. In Section~\ref{muterm} we discuss various 
potential solutions to the $\mu$ problem and in Section~\ref{discussion}
we conclude.

\section{Supersymmetry breaking from a distance}

Our underlying assumption is that all the MSSM matter fields
live on a three brane in extra dimensions whereas the gauge
fields live in the bulk\footnote{We could also have additional larger
dimensions in which only gravity propagates \cite{ADD};
such purely gravitational
dimensions do not alter our framework significantly.}.
Furthermore we assume that supersymmetry is broken
dynamically on a brane which is a distance $d$ away from the
matter brane. The supersymmetry breaking ``source'' brane could
either be a three brane or -- in the case of more than one extra
dimension -- it could also be of higher dimension.
For the explicit calculations in the next section we will
assume that the two branes are at the boundaries of one extra
dimension such that $d=L$.

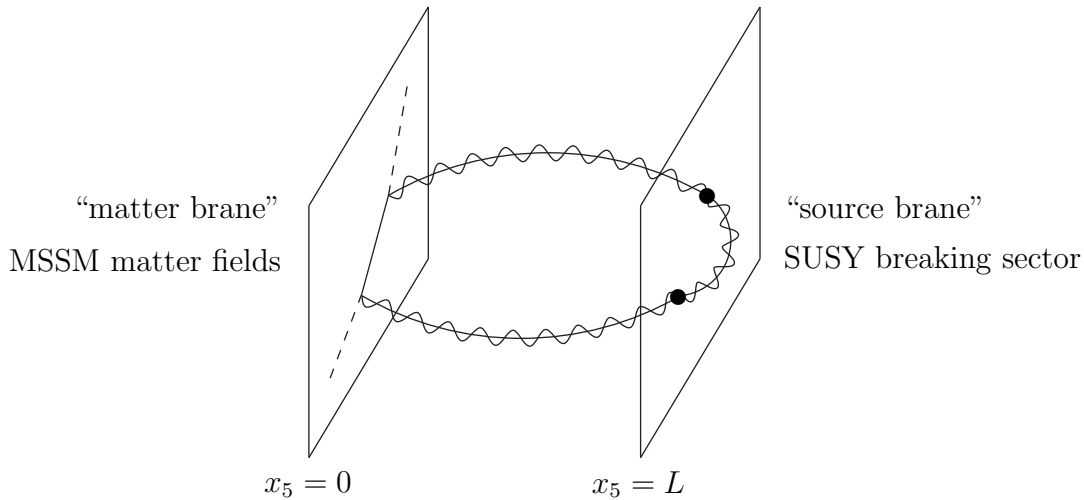
\begin{figure}[!t]
\begin{picture}(420,190)(0,0)
%
%
   \Line( 100, 110 )( 145, 185 )
   \Line( 100,  15 )( 145,  90 )
   \Line( 100,  15 )( 100, 110 )
   \Line( 145,  90 )( 145, 185 )
   \Text( 100,   5 )[c]{$x_5=0$}
   \Text(  90, 110 )[r]{``matter brane''}
   \Text(  90,  90 )[r]{MSSM matter fields}
   \DashLine(  137, 155 )( 130, 114 ){4}
   \CArc(      190,  10 )( 120, 60, 120 )
   \PhotonArc( 190,  10 )( 120, 60, 120 ){3}{10}
   \Line(      130, 114 )( 120, 77 )
   \DashLine(  108,  45 )( 120, 77 ){4}
   \CArc(      180, 180 )( 120, 240, 300 )
   \PhotonArc( 180, 180 )( 120, 240, 300 ){3}{10}
   \CArc(      238,  97 )(  21, 278, 412 )
   \PhotonArc( 238,  97 )(  21, 278, 412 ){3}{5}
   \put( 251, 114 ){\circle*{6}}
   \put( 240,  76 ){\circle*{6}}
   \Line( 225, 110 )( 270, 185 )
   \Line( 225,  15 )( 270,  90 )
   \Line( 225,  15 )( 225, 110 )
   \Line( 270,  90 )( 270, 185 )
   \Text( 225,   5 )[c]{$x_5=L$}
   \Text( 280, 110 )[l]{``source brane''}
   \Text( 280,  90 )[l]{SUSY breaking sector}
\end{picture}
\vskip.1in
\caption{Loop diagram through the bulk, illustrating how scalar 
masses are acquired (and suppressed).}
\label{pretty-picture-fig}
\end{figure}
The basic idea
is that supersymmetry breaking couples directly to the gauginos
in the bulk whereas locality in the extra dimensions forbids
direct couplings between matter fields
and the SUSY breaking sector (see Fig.~\ref{pretty-picture-fig}).

The matter superpartners receive their masses via loop contributions
through the bulk. Depending on the dimensionality of the
bulk and the source brane, as well as the details of the supersymmetry 
breaking sector, their masses are suppressed by varying powers of $d$.
This additional suppression of
the scalar masses relative to the gaugino masses leads to a very
predictive low energy theory: after integrating out the extra
dimensions at the scale $1/L$ we obtain the MSSM with -- to a good
approximation -- only soft SUSY breaking gaugino masses. 

As with gauge mediation and anomaly mediation, this framework solves
the SUSY flavor problem in that the only flavor violation comes from
the Yukawa couplings \cite{sugra,noscale}.\footnote{We are
assuming that the flavor scale,
the scale at which the Yukawa couplings are generated, is at or above the 
compactification scale.} This is because contact terms between MSSM 
matter and the supersymmetry breaking sector are exponentially 
suppressed due to the fact that these are non-local interactions at 
the high scale as in the anomaly mediated scenario of \cite{RSanom}. 
The advantage of our scenario over anomaly mediation is that all
scalar mass squareds (except for the up-type Higgs) receive positive 
contributions from renormalization group running.  Also, in gauge 
mediation, stringent constraints must be imposed on the supersymmetry
breaking sector in order to prevent negative or logarithmically enhanced
scalar masses. Here the scalar masses at the compactification scale
are small enough to render such concerns irrelevant. In addition, 
direct couplings between fields on the source brane and matter brane
are automatically forbidden by locality, while in gauge mediation, 
forbidding messenger-matter couplings requires a non-generic 
superpotential.

Before we go on to describe some specifics of the model, we would like
to discuss a few general properties of the framework.

\noindent{\it \underline{i. strong coupling:}}~
One might worry that our theory is
non-renormalizable and therefore
not predictive. In particular, the gauge coupling in five dimensions
carries dimensions of (mass)$^{-1/2}$ and the theory becomes strongly
coupled at high energies. At lower energies the effects of the strong
coupling are included in the unknown coefficients of higher dimensional
operators. We can estimate the scale of strong coupling $\Mstar$ in 
terms of the volume of the extra dimensions $V$ by using
$${1 \over g_4^2} = {V \over g_{4+n}^2} \sim {V \Mstar^n \over (4 \pi)^2}\ .$$
Here $g_4$ and $g_{4+n}$ are the four- and higher-dimensional
gauge couplings respectively, and we 
defined $\Mstar$ as the scale where the effective dimensionless coupling 
constant is nonperturbative ($g_{4+n} \Mstar^{n/2}\sim 4\pi$).
For example, compactifying on a strip of length $L$ gives
$\Mstar L=16\pi^2/g_4^2\sim \O(100)$.
Thus as long as we only consider external momenta
$\ll \Mstar$ and use $\Mstar$ to cut off loop momenta, 
our effective theory is perturbative and predictive. 

\noindent{\it \underline{ii. mass scales:}}~
The relevant mass scales in our scenario are the compactification scale
$1/L$, the cutoff scale
$\Mstar$ (which -- for simplicity -- we set equal to the scale at which
supersymmetry breaking is communicated\footnote{Messengers
could appear on the source brane at a scale below the cutoff.  
In this case the messenger scale plays the role of the cutoff,
although one must require $\Mstar \gsim 5 L^{-1}$
to suppress higher dimensional contributions to MSSM 
scalar masses.}), and the 
supersymmetry breaking VEV $\sqrt{F}$. $F$ is determined by the scale
at which supersymmetry breaking is mediated and the weak scale by requiring 
that the gaugino masses $m_\lambda$ are of order $M_{weak}$.
As shown in {\it i.}, strong coupling appears at distances about
100 times shorter than $L$, thus $\Mstar\lsim 100\ L^{-1}$.
Therefore only one scale is left undetermined. 
We take the compactification scale to correspond to this parameter 
and allow it to vary between $10^{4}-10^{16}$~GeV\@.  The lower 
limit comes from imposing fine-tuning constraints at the weak scale.
We also impose $L^{-1}\lsim M_{GUT}$ because even higher compactification
scales lead to essentially the same boundary conditions at $M_{GUT}$:
unified gaugino masses and negligible scalar masses.

\noindent{\it \underline{iii. unification and proton decay:}}~
Our framework is fully unifiable, and even though our framework does
not require it we do assume gauge unification. This assumption implies
gaugino mass unification which makes our theory more predictive.
Grand unification might occur at
or below the compactification scale ($M_{GUT}\leq 1/L$) in which case
the running and meeting of the gauge couplings is entirely four-dimensional.
However we could also have $M_{GUT}>1/L$ in which case the couplings
will exhibit power-law running from the compactification scale to the
unification scale \cite{DDG}.  This would lower the GUT scale, 
possibly all the way down to of order $10^{6}$~GeV\@. 
For such low scales proton decay via higher
dimensional operators or $X$ and $Y$ gauge boson interactions represents
a potential disaster. A solution to this problem which would
be very natural in our context is to have quarks and leptons live on
separate ``branes'' in the extra dimensions. The separation forbids
direct local couplings between quarks and leptons, and proton decay 
via $X$ and $Y$ gauge bosons would be exponentially suppressed by the
massive Yukawa propagators of $X$
and $Y$ propagating between the quark- and lepton branes \cite{AS,AHS}.

\noindent{\it \underline{iv. $B\mu$ versus $\tan\beta$:}}~
Naively, our model predicts $B\mu=0$ at
the high scale from which
we can determine $\tan\beta$. However this prediction probably
should not be taken very seriously because, as it stands, the
framework has a $\mu$-problem. The mechanism which sets $\mu$ to
the weak scale will likely also set $B\mu$. Therefore we treat $\tan\beta$ 
as a free parameter in our analysis. We discuss different
attempts at solving the $\mu$ problem in Section~\ref{muterm}. 

To be more specific let us now specialize to the case of one extra
dimension which we parameterize by the coordinate $x_5$. For convenience
we choose the matter and source branes to be located at opposite
ends of the the extra dimension. None of the physics we discuss depends
on this choice, what is important is that the separation is greater than
the short distance cut-off length scale. Coupling supersymmetric
three branes to a supersymmetric bulk gauge theory is complicated by
the fact that the minimal amount of supersymmetry in five dimensions
corresponds to $N=2$ supersymmetry in four dimensions. Ignoring auxiliary
fields the minimal five-dimensional vector superfield contains a
real scalar $\phi$, a vector $A_N$, and a four component spinor $\lambda$.
They decompose as follows when reduced to four dimensions
\bea
(\phi \quad A_N \quad \lambda) \quad &\longrightarrow& \quad
(A_\mu \quad \lambda_L) \quad + \quad (\phi + iA_5 \quad \lambda_R) \cr
{\rm 5-d\ \ vector}\hskip.25in &&\quad{\rm 4-d\ \  vector}
\hskip.7in {\rm 4-d \ \  chiral}
\eea
where $\lambda_{R/L}\equiv \frac12(1\pm\ \gamma^5) \lambda$.
In order to break the additional supersymmetry and to give
mass to the unwanted adjoint chiral superfield we compactify
the fifth dimension on an orbifold. We choose a $Z_2$ orbifold
which acts as $x_5 \rightarrow -x_5$ on the circle $x_5\in (-L,L]$. The
$Z_2$ breaks half of the supersymmetries by distinguishing the components
of the vector superfield. We take it to act as
\bea
(A_\mu \quad \lambda_L)\ (x,x_5)\quad &\longrightarrow\quad\quad& \quad
      (A_\mu \quad \lambda_L)\ (x,-x_5) \cr
(\phi + iA_5 \quad \lambda_R)\ (x,x_5)\quad&\longrightarrow \quad -&\ 
(\phi + iA_5 \quad \lambda_R)\ (x,-x_5) \ ,
\label{orbi}
\eea
which allows a massless mode for the 4-d vector but not for the
4-d chiral superfield. In practice this means that we expand the
fields of the vector superfield with cosine KK wave functions,
whereas the chiral superfield is expanded in sine modes\footnote{For
a more detailed description of the orbifold we refer
the reader to Ref.~\cite{MP}.}.

In order to write couplings between the bulk fields and brane
fields we note that at the boundaries the components of the $N=2$ fields
which are non-vanishing exactly correspond
to a 4-d vector multiplet. Therefore, we can couple them to 
boundary fields in the same way as we would couple a four-dimensional
$N=1$ vector multiplet. The action is then 
\beq
\L=\int d^5x [ \ \L_5\ +\ \delta(x_5)\ \L_m\ +\ \delta(x_5-L)\ \L_s\ ] 
\label{lagrangian}
\eeq
where $\L_5$ is the bulk Lagrangian for the SM gauge fields
\beq
\L_5= - \frac12 \, {\rm tr} \, (F_{MN})^2 
      + \, {\rm tr} \, (\overline\lambda i \Gamma^M D_M \lambda) 
      + \dots 
\eeq
Here $M,N$ label all five dimensions, $\Gamma^\mu\equiv \gamma^\mu$
are the usual four-dimensional gamma matrices, $\Gamma^5=i\gamma^5$,
$D_M$ is the five-dimensional covariant derivative, and
we suppressed all terms involving the scalar adjoint $\phi$
and auxiliary fields. 
Note that there is such an expression for each
of the gauge multiplets in $SU(3)\times SU(2) \times U(1)$. 

The supersymmetry breaking sector on the source brane at $x_5=L$
can be quite arbitrary. It is one of the strengths of our framework
that it is compatible with many different SUSY breaking sectors.
The only requirement of this sector is
that the gaugino masses generated are not highly suppressed compared
to the scale $F/\Mstar$.  If there is a singlet chiral superfield $S$
with an $F$ at or near the supersymmetry breaking scale
squared, then it will give the dominant contribution to gaugino masses.
Though it is possible to produce a viable spectrum
even without a singlet, we will assume the singlet exists.  We briefly
discuss the alternative in Section~\ref{discussion}.

\subsection{Source brane action}

The source brane action is in general very complicated and involves
all the fields required to break supersymmetry dynamically as well
as couplings to the bulk gauge fields. However, in order to compute
the MSSM gaugino and scalar masses only a small subset of the
operators are necessary. If we assume that the leading
supersymmetry breaking VEV is the $F$ component of a singlet
chiral superfield $S$, then we only need terms of the effective action
which couple this singlet to the MSSM gauge fields. 
The leading superpotential term which couples $S$
to the bulk gauge fields and which contains only two field
strengths $W$ is of the form
\beq
\L_s  \sim \int d^2 \theta\ {S\over \Mstar^2}\  W  W \ + h.c.  
\label{gaugmass1}
\eeq 
The gauge field strength superfields $W$ here are five-dimensional
with mass dimension two, and the $S$ field is four-dimensional with
mass dimension one. This term
contributes a gluino mass $\delta(x_5-L)\ F_S/\Mstar^2$ which is
localized on the source brane.
Terms with more powers of $S$  do not give rise to new supersymmetry
breaking interactions; they only give 
higher order (in $S/\Mstar$) contributions to the gluino mass and
are therefore irrelevant.
Next we consider the most general supersymmetry breaking K\"{a}hler
potential terms with only two $W$s, arbitrary powers of $S$
and no derivatives. (Note that Lorentz invariance forbids terms
of the form $W\overline W$.) 
The leading non-vanishing terms contain a single
$S^{\dagger}$
\beq
\L_s  \sim \int d^2 \theta d^2\overline\theta \ 
{1\over \Mstar^3}\ S^\dagger W W
(1+\frac{S}{\Mstar}+\cdots ) = \int d^2 \theta \ 
{F^\dagger_S\over \Mstar^3}\ W W 
(1+\frac{S}{\Mstar}+\cdots )  \ .
\eeq 
Equivalent terms with less suppression are already contained in the
superpotential. Therefore there are no important supersymmetry breaking
terms in the K\"{a}hler potential with no derivatives.

Using arguments similar to those given above and the constraint
$\overline D_{\dot{\alpha}} \overline W^{\dot{\alpha}}= D^\alpha W_\alpha$
it is straightforward to determine all K\"{a}hler
potential terms with two derivatives which give rise to new supersymmetry
breaking. The most important such terms are non-supersymmetric
contributions to kinetic terms such as
\beq
\L_s  \sim \int d^2 \theta d^2\overline\theta \ 
{S^\dagger S\over \Mstar^5}\  W D^2 W \longrightarrow 
{F_S F^\dagger_S \over \Mstar^5}\ \overline \lambda_L \delslash \lambda_L \ .
\label{kahler}
\eeq 
In the next section we will see that this supersymmetry breaking correction
to the gaugino kinetic term gives rise to
(small) scalar masses when inserted into loop diagrams.

\section{The MSSM scalar and gaugino masses}
\label{loops}

In this section we compute the MSSM soft supersymmetry breaking
masses that result from integrating out the extra dimensions. We
always assume that loop momenta are larger than $L^{-1}$. Smaller
loop momenta are more conveniently dealt with 
by considering the four-dimensional effective theory, as we do
in Section~\ref{pheno}.

It is straightforward to determine the gaugino masses resulting
from the term eq.~(\ref{gaugmass1}) on the source brane
by expanding the five-dimensional
gaugino fields in KK modes. The zero mode which corresponds to the
light four-dimensional gaugino has an $x_5$-independent wave
function, which when normalized to produce a canonical kinetic term
has height $1/\sqrt{L}$. Thus the gaugino mass is
\beq
m_\lambda={1\over \Mstar L}\ {F_S\over \Mstar} \ .
\eeq

To calculate the scalar masses more effort is required. The leading
contributions come from loop diagrams which involve both the
scalars on the matter brane as well as supersymmetry violating
operators on the source brane (Fig.~\ref{pretty-picture-fig}). 
Any of the fields in the five-dimensional 
gauge multiplets can be exchanged. In principle, this leads 
to a large number of diagrams
which need to be calculated. However, since we are only interested
in showing that the scalar masses are small, we only compute two
representative diagrams with bulk fermion exchange. The other
diagrams are of comparable size and therefore also negligible.

It is most convenient
to compute the five-dimensional Feynman diagrams in momentum space
in four dimensions and position space in the fifth dimension.
This mixed position-momentum space calculation is well adapted to 
the symmetries of the problem (translation invariance in four 
dimensions but broken translation invariance in $x_5$). The necessary
propagators are obtained by partially Fourier transforming normal
momentum space propagators, whereby care needs to be taken to properly
take into account the orbifold boundary conditions eq.~(\ref{orbi}).
For example, the scalar $\phi$ propagator with 4-d Euclidean
momentum $q^2$ propagating from coordinate $b$ to $a$
in the $x_5$-direction is
\beq
\P_0(q^2;b,a)=\frac2L \sum_{n,m=1}^\infty
         \sin(-p_n a){\delta_{nm} \over q^2+p_n^2} \sin(p_m b)
        \sim -\int_{-\infty}^{\infty} \frac{dp}{2\pi}
        {e^{ip(b-a)} \over q^2+p^2}
        = -{e^{-|b-a|\sqrt{q^2}} \over 2\sqrt{q^2}} \ .
\label{sprop}
\eeq
We have implemented the orbifold boundary conditions for $\phi$ by 
expanding in sine modes with Fourier momenta $p_n=n \pi/L$.
By approximating the sum with an integral
we have assumed large volume ($L> 1/\sqrt{q^2}$).
Performing the sum exactly is straightforward \cite{AGS} but 
not necessary for our purposes.

Analogously the fermionic propagator is obtained by Fourier expanding
the momentum space propagator in sine and cosine wave functions for
the right and left handed components respectively
\beq
\P(q;a,b)=\frac2L \sum_{n,m=0}^{\infty}
\left[P_L {\cos(p_n a) \over \sqrt{2}^{\delta_{n0}}}-P_R \sin(p_n a)\right]
\delta_{nm}
{\qslash-i\gamma_5 p_n \over q^2+p_n^2}
\left[P_R {\cos(p_m b) \over \sqrt{2}^{\delta_{m0}}}+P_L \sin(p_m b) \right]
\ .
\eeq
Again $p_{n}=n \pi / L$, the factor of $\sqrt{2}^{\delta_{n0}}$
arises from the different wave function normalization of the zero mode,
and again we have Wick-rotated the four-momentum to Euclidean space.
At the boundaries $x_5=0$ and $x_5=L$ only the left-handed gaugino component
is non-vanishing and can be coupled
directly to the scalars and the supersymmetry breaking sector.
The other components
require $\del_5/\Mstar$ derivatives in the couplings and are therefore
subleading (after regularization and renormalization of the
divergent momentum integrals).  
We only keep the leading cosine components of the propagator.
Summing over momenta we find
\beq
\P(q;0,L)={P_L \qslash\  \over q \sinh(qL)}
      \sim {2 P_L \qslash\ \over q} e^{-qL} \ .
\eeq
Armed with this very simple formula for the 5-d gaugino propagator it
is straightforward to compute the diagram with two gluino mass insertions
in Fig.~\ref{pretty-picture-fig}.
Ignoring Casimirs and factors of 2 we find
\bea
g_5^2 \left(\frac{F_S}{\Mstar^2}\right)^2&\times& \int {d^4q \over (2 \pi)^4}
 \ \tr \left[{1\over \qslash}\ P_L \P(q;0,L)\ C\ \P^T(q;L,L)\ 
      C^{-1}\ \P(q;L,0)\right] \cr
 &\sim& {g_5^2 \over 16 \pi^2} \left(\frac{F_S}{\Mstar^2}\right)^2 {1\over L^3}
      = {g_4^2 \over 16 \pi^2} m_\lambda^2 \ .
\label{fsm}
\eea
We see that the scalar masses are suppressed by three powers of the brane
separation which can be absorbed into the four-dimensional gauge couplings
and gaugino masses. Thus we find
that the scalar mass contributions from this diagram are smaller
compared to the gluino masses by a loop factor. Note that these small
contributions to the scalar masses are flavor universal and do not
give rise to flavor changing effects.

As a second example we compute the contribution from a supersymmetry
breaking gaugino wave function renormalization insertion eq.~(\ref{kahler})
on the source brane. We find
\bea
g_5^2\ \frac{F_S^2}{\Mstar^3}&\times& \int {d^4q \over (2 \pi)^4}
   \ \tr \left[{1\over \qslash}\ P_L\ \P(q;0,L)\ \qslash
          \ \P(q;L,0)\right] \cr
 &\sim& {g_5^2 \over 16 \pi^2} \left(\frac{F_S}{\Mstar^2}\right)^2 
        {1\over \Mstar L^4}
      = {g_4^2 \over 16 \pi^2} m_\lambda^2 {1\over \Mstar L}\ ,
\label{ssm}
\eea
which is suppressed by an additional power of the separation compared
to the contribution of eq.~(3.5).  Note that one could have obtained this 
result from dimensional analysis: soft scalar masses require two insertions 
of supersymmetry breaking $F_S^2$, the powers of $\Mstar$ in the denominator 
are determined by the dimensionality of the operators which we 
insert on the source brane, the exponent of the
separation $L$ can then be determined by dimensional analysis.
This dimensional analysis also shows that diagrams involving
even higher dimensional operators (such as operators with additional
$\del_5/\Mstar$ derivatives) are suppressed by additional powers
of $(\Mstar L)^{-1}$.

In summary we find that the MSSM scalar mass squareds are suppressed relative
to the gaugino masses by at least a loop factor, and are therefore
negligible compared to the masses which are generated from the
(four-dimensional) renormalization group evolution between the
compactification scale and the weak scale. This conclusion
also holds for the other soft supersymmetry breaking parameters involving
matter fields, the $A$-terms and $B\mu$. Note that these contributions
to soft parameters are flavor-diagonal and are thus irrelevant with
regards to bounds on FCNCs.

\subsection{Example: gauge mediation with branes}

Here we demonstrate the above results with an explicit model for
the supersymmetry breaking sector on the source brane. We take the
source brane action to be identical to the ordinary messenger sector
of gauge mediation where the SM gauge
fields are replaced by the boundary values of the bulk gauge fields
\beq
\L_s  = \int d^4\theta \ Q^\dagger e^{2gV[A^\mu,\lambda_L]}Q + 
      \widetilde Q^\dagger e^{-2gV[A^\mu,\lambda_L]}\widetilde Q      
     + \int d^2 \theta\ S Q \widetilde Q \ .
\eeq 
Here $Q+\widetilde Q$ are the messenger chiral superfields
which we take to transform under the SM gauge interactions with the
quantum numbers $5+\overline5$ of $SU(5)$. The vector
superfield $V[A^\mu,\lambda_L]$ contains the SM gauge fields and
gauginos in the normalization appropriate for five-dimensional fields.
The $S$ field has been rescaled to absorb the Yukawa coupling,
and as in ordinary gauge mediation we assume that it
acquires supersymmetry preserving and violating expectation values
$$S=\Mmess+F_S\theta^2 \ .$$
Then the messenger fermions obtain the Dirac mass $M$
whereas the messenger scalars in
$Q$ and $\widetilde Q$ acquire the (mass)$^2$ $\Mmess^2\pm F_S$.
Note that role of the cut-off (or new physics) scale in our more general
effective theory of the source brane is played by the messenger mass in this
example.

The bulk gauginos obtain a mass which is localized on the source brane from
a one-loop diagram with messenger scalars and fermions in the loop as
in ordinary gauge mediation. Since the messengers
are stuck to the brane this calculation is entirely four-dimensional
and we find the effective gaugino mass
\beq
{g_5^2 \over 16 \pi^2 L}\ \frac{F_S}{\Mmess}  \ .
\eeq
The gauge couplings $g_5$ here
are five-dimensional (and in the GUT normalization); they
are related to four-dimensional couplings by $g_5^2/L=g_4^2$. 
We see that our gaugino masses are identical to ordinary four 
dimensional gauge mediation gaugino masses.

The computation of the scalar masses is more involved. We simply quote
the result obtained by Mirabelli and Peskin \cite{MP} who 
computed the scalar masses at two loops for arbitrary separation. 
Expanding to second order in $F_S$ and to leading order 
in $(L \Mmess)^{-1}<1$ their result reduces to
\beq
m_5^2=2 C \left({g_5^2 \over 16 \pi^2 L}\frac{F_S}{\Mmess}\right)^2
      {\zeta(3) \over (\Mmess L)^2} = m_4^2 {\zeta(3) \over (\Mmess L)^2} \ .
\label{ginomichael}
\eeq
Here $m_5^2$ is the scalar mass in the five-dimensional theory,
$m_4^2$ is the ordinary four-dimensional gauge mediation result,
and $C$ is a group theory factor of order one which depends on the quantum
numbers of the matter and source fields.
The important conclusion is that the scalar mass squareds are suppressed
relative to gaugino masses by a factor of $1/(\Mmess L)^2$. Again assuming
a distance which is at least a factor of 5 larger than the messenger scale,
we find that eq.~(\ref{ginomichael}) is negligible compared to the
masses which are generated from four-dimensional running.

To compare this to our general analysis of the previous section note that
the scalar mass squared scales as $1/L^4$ when expressed in terms of
five-dimensional quantities. This is in agreement with the scaling
found for eq.~(\ref{ssm}). The scalar mass contribution scaling as $1/L^3$
eq.~(\ref{fsm}) corresponds to a three loop diagram in the gauge
mediation model. We see that the scalar mass squareds are suppressed 
by at least $(\Mmess L)^{-2}$ or a loop factor.

\section{Spectrum and phenomenology}
\label{pheno}

To calculate the spectrum in our scenario we use the renormalization 
group to connect the physics near the cutoff scale with the weak scale.  
In particular, there are two scaling regions that must be considered 
when evolving masses and couplings:  between the cutoff scale and the 
compactification scale, and between the compactification scale and 
the weak scale.  

Above the compactification scale the theory is five-dimensional and we
need to evolve masses and couplings according to five-dimensional evolution
equations.  Happily this turns out to be rather straightforward.  The 
calculations of the previous section showed that the scalar masses which 
are generated above the compactification scale are negligible.
Therefore we do not need to evolve either scalar masses or $A$-terms in 
the five-dimensional theory. 
The gaugino mass evolution is important however.  For this purpose it
is most convenient to think of the theory as four-dimensional with KK
excitations.  Across each KK threshold, the four-dimensional gauge and
gaugino beta functions are modified, and such corrections must be included
to calculate the low energy spectra.  However, the ratio of 
the gaugino mass to the gauge coupling squared is invariant to one-loop, 
as in the normal four-dimensional case\footnote{This can also be seen by 
noting that the gaugino mass is in the same supermultiplet as the 
holomorphic gauge coupling $\tau$ and therefore evolves in parallel. 
For the usual 4-d arguments \cite{GR} to go through even in 
the theory with KK modes, the orbifold boundary conditions must preserve 
4-d $N=1$ supersymmetry as in our framework.}
(for discussion of this, see Ref. \cite{Kribs-gaugino}).
Summing over the towers of KK thresholds up to the cutoff $M_*$ can be
represented by terms that resemble corrections to the renormalization
group equations to both the gauge couplings \cite{DDG} and gaugino 
masses and gives the same result \cite{KKMZ}.  
Specifically, assuming gauge coupling unification, the relations
\beq
\frac{M_1}{g_1^2}\;=\;\frac{M_2}{g_2^2}\;=\;{M_3 \over g_3^2}
\label{GUT-relation-eq}
\eeq
hold at any scale to one-loop order in the $\beta$-functions.
This means that we can incorporate the extra dimensional running of the
gaugino masses simply by starting with the boundary condition,
Eq.~(\ref{GUT-relation-eq}), at the compactification scale.
Note that this relation also implies that our predictions for
gaugino masses will be nearly independent of the compactification scale.

Just below the compactification scale, our theory is four-dimensional
with nonzero gaugino masses, vanishing scalar masses, and vanishing 
trilinear scalar couplings.  Scalar masses and trilinear scalar 
couplings are regenerated through renormalization group evolution 
between the compactification scale and the weak scale, and this provides 
the basis to calculate the spectrum and phenomenology.  The parameters 
for the model can be chosen to be
\begin{eqnarray*}
L^{-1}, \; M_{1/2}, \; \tan\beta, \; {\rm sign}(\mu) \; .
\end{eqnarray*}
Here $M_{1/2}$ is the common gaugino mass at the unification scale.
For $L^{-1}<10^{16}$ GeV the individual gaugino masses at the
compactification scale can be determined from
$M_a(L^{-1})/g^2_a(L^{-1})=M_{1/2}/g^2_{\rm unif}$
with $g_{\rm unif}\sim 0.7$.
Imposing electroweak symmetry breaking constraints at the 
weak scale determines $\mu^2$, leaving $\tan\beta$ and 
${\rm sign}(\mu)$ unknown\footnote{While $B\mu$ is what appears in
the Lagrangian, we choose to parameterize our ignorance by $\tan\beta$.}. 
Generally, the scalar masses are proportional 
to $M_{1/2}$ to reasonable accuracy unless Yukawa coupling effects
are large (i.e., particularly for the up-type Higgs mass), or weak 
interaction eigenstate mixing is important (i.e., stau masses at moderate
to large $\tan\beta$).
\begin{figure}[t]
\centering
\epsfxsize=5.6in
\hspace*{0in}
\epsffile{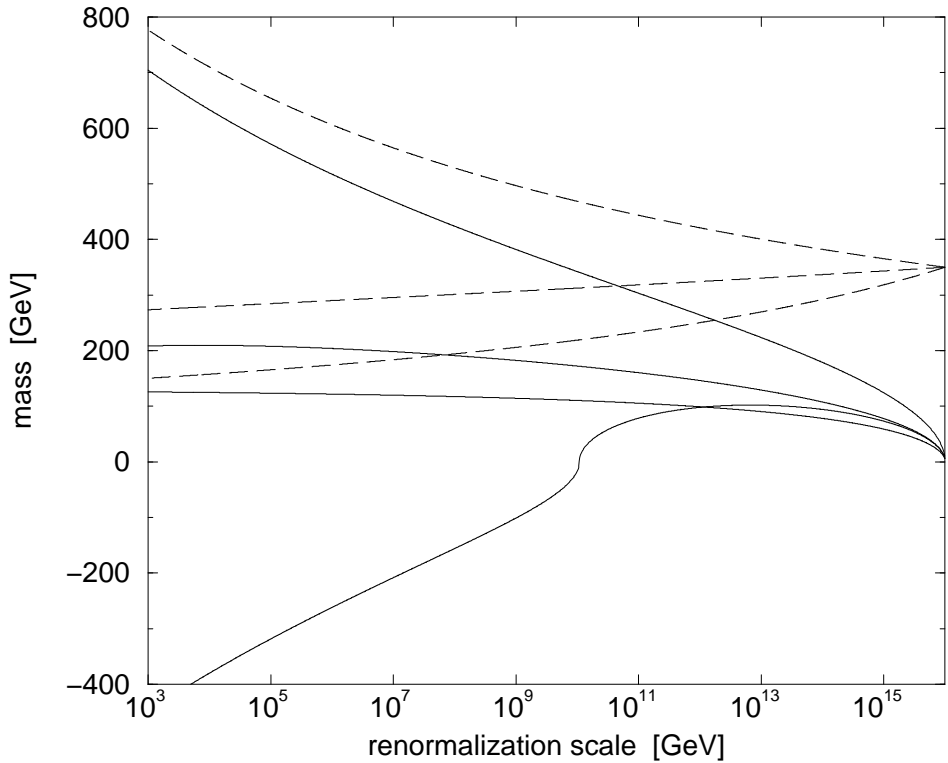}
\caption{Evolution of several soft masses as a function of the
renormalization scale with the input parameters $L^{-1} = 10^{16}$ GeV, 
$M_{1/2} = 350$ GeV, and $\tan\beta = 10$.  The (top, middle, bottom) 
dashed lines correspond to ($M_3$, $M_2$, $M_1$), while the solid lines 
from top to bottom correspond to $m_{\tilde{Q}_1}$, $m_{H_d}$, 
$m_{\tilde{\tau}_1}$, ${\rm sign}(m_{H_u}^2) |m_{H_u}^2|^{1/2}$
respectively.  (The kink in the up-type Higgs mass is due to taking 
the square-root.)}
\label{example-fig}
\end{figure}
As a first example, we take $L^{-1} = 10^{16}$ GeV, $M_{1/2} = 350$ GeV, 
and $\tan\beta = 10$, and show in Fig.~\ref{example-fig} the evolution 
of the soft masses as a function of the renormalization scale. 
Several generic features are evident from the graph:
Gaugino masses evolve in parallel with gauge couplings;
the ratios $M_3/M_1$ and $M_3/M_2$ increase 
as the renormalization scale is decreased, causing larger squark 
masses relative to slepton and Higgs masses.  Initially, $m_{H_u}^2$ 
runs toward positive values, but is quickly overcome
by interactions with the heavy stops and runs to negative values
at the weak scale.  With these parameters, the stau is the lightest sparticle 
of the MSSM spectrum.

\begin{figure}[b]
\centering
\epsfxsize=5.6in
\hspace*{0in}
\epsffile{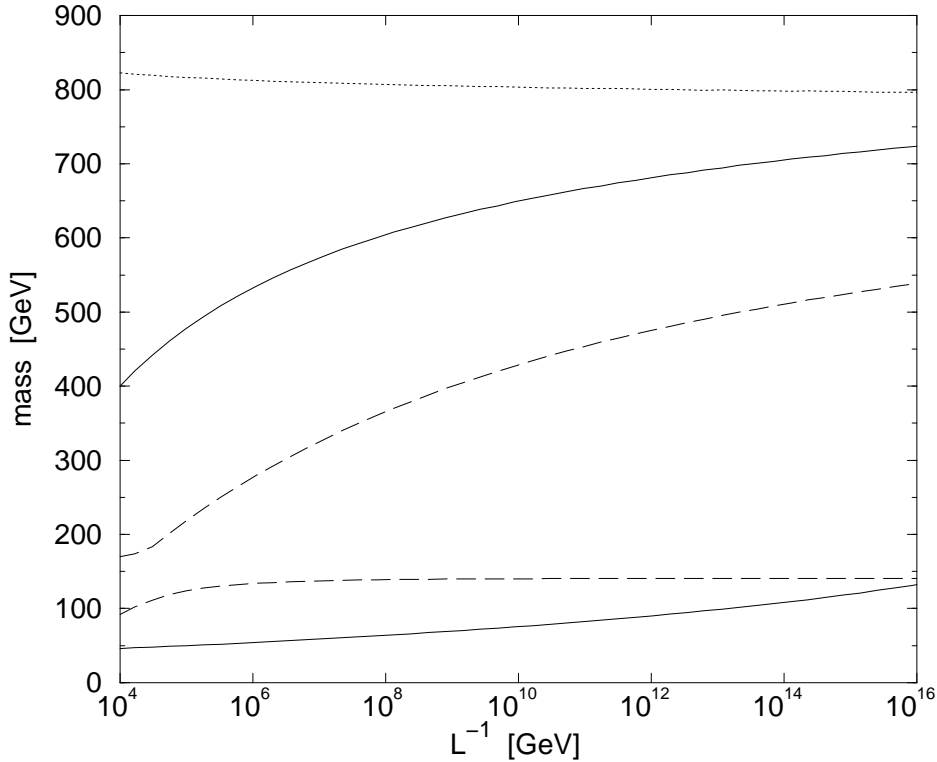}
\caption{The weak scale masses for several sparticles
are shown as a function of the compactification scale $L^{-1}$
with $M_{1/2} = 500$ GeV and $\tan\beta = 3$.
The top dotted line is $m_{\tilde{g}}$, the top and bottom solid
lines are $m_{\tilde{u}_L}$ and $m_{\tilde{\tau}_1}$, and the
top and bottom dashed lines are $m_{\tilde{N}_3}$ and $m_{\tilde{N_1}}$.
We emphasize that $L^{-1}$ is parameter of our model not
to be confused with the renormalization scale.}
\label{compact-fig}
\end{figure}
The results of the previous analysis are the same as those in ``no-scale'' 
supergravity models \cite{noscale}.  However, in our framework, the detailed 
phenomenology depends on the compactification 
scale.  Obviously the size of the scalar masses depends
on the extent of evolution, proportional to $\sim M_{1/2}^2 \log(M_Z L)$,
but also derived parameters such as $\mu$ are sensitive to
the compactification scale.  In Fig.~\ref{compact-fig} 
we show the weak scale masses of several MSSM fields as a function 
of the compactification scale for $M_{1/2} = 500$ GeV and $\tan\beta = 3$. 
A generic prediction of our model is
that the stau is the NLSP for most compactification scales.  
However, we note that for very large 
$L^{-1} \gsim 10^{16}$~GeV with small $\tan\beta \lsim 3$, the 
lightest neutralino $\tilde{N}_1$ becomes the NLSP (or LSP, as
discussed below).
The kinks in the mass contours of $\tilde{N_1}$ and $\tilde{N}_3$ in
Fig.~\ref{compact-fig} indicate a ``cross over'' in the dominant
interaction eigenstate content of the neutralinos from bino-like 
to Higgsino-like as the compactification scale is lowered 
below $L^{-1} \sim 10^{5}$ GeV.  This suggests that, for example, 
a measurement of the gauge eigenstate content of the lightest neutralino 
is sensitive to the compactification scale.

\begin{figure}[t]
\centering
\epsfxsize=5.6in
\hspace*{0in}
\epsffile{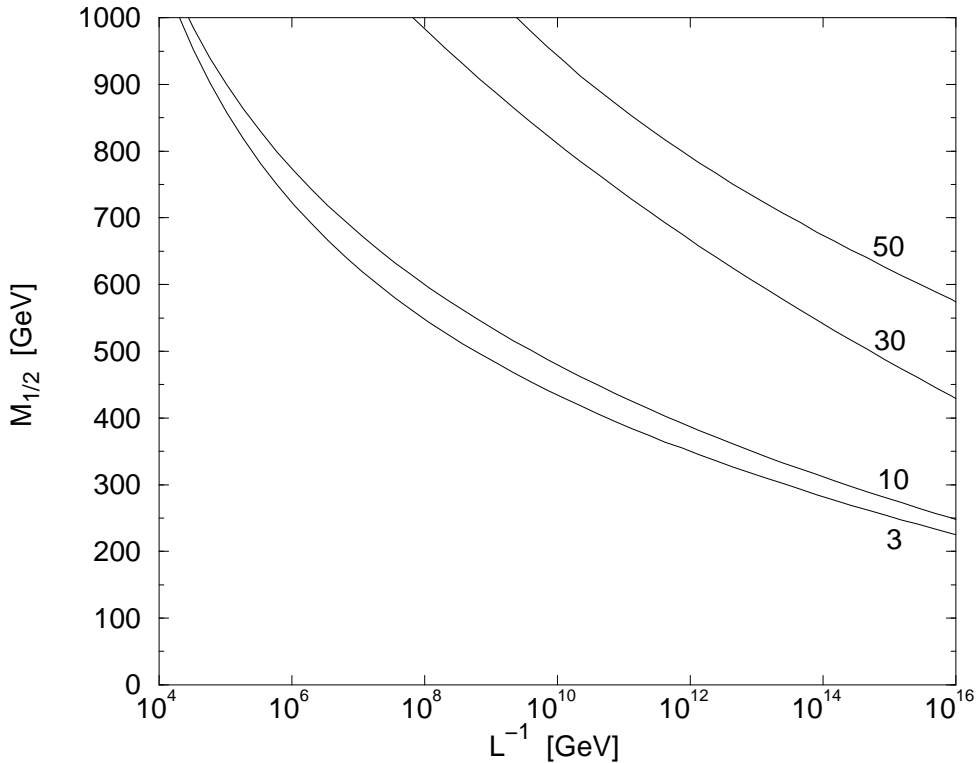}
\caption{The lower bound on $M_{1/2}$ as a function of the
compactification scale obtained by requiring that all charged sparticles
and the lightest Higgs are heavier than the current LEP limit 
(of about $90$ GeV).  The contours correspond to the limits for
particular values of $\tan\beta$.}
\label{Mhalf-bound-fig}
\end{figure}

The scaling of scalar masses proportional to $\sim M_{1/2}^2 \log(M_Z L)$
is clearly visible from Fig.~\ref{compact-fig}; it affects the squarks 
most dramatically but is also important for sleptons, particularly the
lightest (mostly right-handed) stau. 
Note that this allows us to extract significant limits on $M_{1/2}$ as a
function of $L^{-1}$ by requiring that the stau avoids the lower bounds
from the recent LEP searches for charged sparticles.
In Fig.~\ref{Mhalf-bound-fig} we show the lower bound on $M_{1/2}$
as a function of the compactification scale.  The best bound comes
from the lower limit on the stau mass, although low $\tan\beta \lsim 3$
is also restricted by the limit on the lightest Higgs boson.
In addition, notice that for large values of $\tan\beta$, the
lower bound on $M_{1/2}$ is considerably strengthened. This is due to
large mixing in the stau mass matrix from the off-diagonal 
term proportional to $m_{\tau} \mu \tan\beta$ that reduces the
mass of the lightest stau mass eigenstate.

As we implied above, the gravitino is the LSP for most of the parameter
space. Assuming that $F_S$ is the largest supersymmetry breaking VEV, its
mass is given by $m_{3/2}\sim F_S/M_{\rm Planck}$. However, for very large
compactification scales the mass of the stau which roughly scales as
$M_{1/2}\sim F/(\Mstar^2L)$ can become smaller than $m_{3/2}$. 
Then the stau could become the LSP which is probably in conflict 
with cosmology.  The turn over occurs when 
$M_{\rm Planck}\sim F/(\Mstar^2 L)$ or $L^{-1} \sim 10^{14-16}$ GeV\@. 
We find it amusing that coincidentally the largest
compactification scales also correspond to the regime where the lightest
neutralino can be LSP, which would render the stau cosmologically safe again.
The viability of this regime clearly deserves further study.

Superpartner production at colliders always results in two or more 
NLSPs (directly or indirectly),
each of which then decays into the LSP with a decay length that
is expected to be at least of order the size of the detector.
If the stau is the NLSP one expects clearly visible charged
stau tracks in detectors resulting from meta-stable staus
that escape the detector\footnote{A stau NLSP could also have interesting
interesting implications for cosmology \cite{cosmo}.}.
Strategies to extract this signal from the muon background
have been explored in Ref.~\cite{stau-track-signals}, with
the result that rather significant regions of parameter space
can be probed.  For very small compactification scales
$L^{-1} \lsim 10^5$ GeV, it is possible that the stau decay length 
could be measurable.  In the small 
region of parameter space where the neutralino is the (N)LSP, the 
characteristic signal is missing energy, analogous to gauge-mediation 
models with a large messenger scale, or ordinary supergravity models.

\section{The $\mu$ term}
\label{muterm}

As in other models of supersymmetry breaking we appear to have a
$\mu$ problem in our framework.  The $\mu$ term is the 
dimensionful superpotential coupling of $H_u$ and $H_d$ and is required to
be at the weak scale in order to naturally produce electroweak symmetry
breaking while maintaining agreement with experimental lower bounds on
sparticle masses.

{}From naturalness \cite{thooft}, one would expect a dimensionful quantity
to be of order the fundamental scale in the model, in our case $\Mstar$.
However, it is well known that superpotential couplings can easily be
non-generic, and $\mu$ can also be set to zero by imposing a discrete
version of a Peccei-Quinn symmetry \cite{pq}. 
Allowing the discrete symmetry to break
spontaneously with the breaking of supersymmetry, it is easy to produce
a weak-scale $\mu$ term.  However, it is difficult to
produce soft Higgs-mass terms at the same scale (they normally come out
too large).  Here we present some possible solutions to the $\mu$ problem.  
This new framework may allow for more novel solutions and we leave these
for future work.

Perhaps the most elegant possibility for a solution lies with the 
Next-to-Minimal Supersymmetric Standard Model (NMSSM) \cite{nmssm}.  
Inserting this mechanism into our framework means adding 
a gauge singlet $N$ to the matter brane and replacing the $\mu$ term in 
the superpotential by:
\beq
W_N = \lambda N H_u H_d + {k\over 3} N^3 .
\eeq
As the soft masses are run from the compactification scale to the weak
scale, $N$ develops a scalar vacuum expectation value of order the weak
scale for some range of parameters $\lambda$ and $k$.  Thus an effective
$\mu$ term is produced.  This mechanism was thoroughly analyzed by
de Gouv\^ea, Friedland and Murayama in the context of gauge mediation with
a range of messenger scales \cite{dGFM}.  They found the NMSSM could
produce a $\mu$ term but only at the expense of giving unacceptably light
masses to Higgs bosons and/or sleptons.  However, our boundary conditions
are different and may push the results in the right direction.

A twist on this solution is to put the singlet $N$ in the bulk.  The first
obvious requirement is that the $F$ term of $N$ must be suppressed relative
to the supersymmetry breaking scale.  Otherwise, $F_N$ would generically
give non-universal scalar masses.  If there are
fields on the source brane charged under the SM gauge group 
(say, extra vector-like quarks) to which $N$ couples, then a solution 
may be found as suggested in \cite{AgGr,dGFM}.  This solution appears to be 
fine-tuned and the fine tuning comes from the same source as the fine tuning 
in the MSSM.  So this mechanism could explain the dynamical origin of the 
$\mu$ term, but it does not give a dynamical reason for the cancellation 
of large soft parameters.  One could also consider more than one singlet
and could place singlets in the bulk or on either of the branes.
This would allow certain couplings to be small or vanish, possibly giving
the right parameter values for a natural $\mu$ term as in \cite{han}.

The suggestion of Chacko et al. \cite{CLMP} to put the Higgs fields in 
the bulk (while keeping $S$ on the source brane) is also interesting.
The $\mu$ term could be produced on the 
opposite brane via the Giudice-Masiero mechanism \cite{GiudMas}.  The 
operators in the (5-dimensional) Lagrangian would be:
\beq
\int d^4\theta \left[\lambda_{\mu} \frac{S^{\dagger}}{\Mstar^2}H_u H_d
	+ \lambda_B \frac{S S^{\dagger}}{\Mstar^3} H_u H_d +
        \frac{S S^{\dagger}}{\Mstar^3} (\lambda_{uu} H_u H_u^{\dagger}
	+ \lambda_{dd} H_d H_d^{\dagger})+ h.c.\right] \delta(x_5-L)
\eeq
where the coupling constants $\lambda_i$ are dimensionless.
Thus the natural value of $\mu$ would be $F_S/(\Mstar^2 L)$,
where as the natural value of the soft parameters $B \mu$, $m_{H_u}^2$
and $m_{H_d}^2$ would be $F_S^2/(\Mstar^3 L) \sim \mu (F_S/\Mstar)$.  
We find the standard problem of producing soft terms which are too large.  
We could of course set the appropriate couplings to be small 
($\sim (\Mstar L)^{-1}$), however we do not have a compelling 
theoretical reason for doing so.  Also we note that placing the
Higgs fields in the bulk changes the spectrum of the model
significantly as their scalar masses would be generated above
the compactification scale.  We have found the resulting phenomenology
is viable and thus a detailed analysis would be interesting.

In summary, there exist a number of ways to produce a $\mu$ term 
dynamically in our scenario.  However, they all appear to require
small or fine-tuned parameters.  Thus finding a natural origin for
$\mu$ and B$\mu$ of the right size is still an open problem.

\section{Discussion}
\label{discussion}

We have presented a model of supersymmetry breaking in extra dimensions
in which only gauginos receive soft masses at a high scale, and scalar
masses come dominantly from renormalization group running.  The model
clearly avoids the supersymmetric flavor problem, and all scalar mass squareds
(except for a Higgs) are positive at the weak scale.  The model is
highly predictive, depending only on three parameters and a sign 
($M_{1/2}$, $L^{-1}$, $\tan \beta$, and the sign of $\mu$), 
and allows for compactification scales as low as $10^4$~GeV\@.

For simplicity, we required the gaugino masses to unify at or above
the compactification scale.  This comes from the assumption that the
theory is unified at a high scale and that threshold effects are small.
By relaxing either assumption, one could impose more general boundary
conditions, i.e., with split gaugino masses.  As long as the gluino
is heavy enough to give squark masses larger than Higgs masses,
and the bound in Fig.~\ref{Mhalf-bound-fig} is respected
(properly reinterpreted as a bound on $M_1$), then supersymmetry
breaking through transparent extra dimensions would still work
perfectly.

The only requirement on the source brane is that there exists a
singlet whose $F$ component is comparable to the scale of supersymmetry
breaking.  However, even this requirement may be relaxed.  Without
a singlet, the main contribution to the gaugino masses is via anomaly
mediation -- a one-loop effect \cite{RSanom,GLMR}.  The dominant
contributions to the scalars would come from the anomaly-mediated
contributions and from non-renormalizable operators inserted in loops 
(as in Sec.~\ref{loops}), both of which are flavor-blind.  For small
values of $\Mstar L$, the latter may dominate allowing for a 
(different) realistic spectrum.

While the size of the compactification scale does not allow for direct
detection of KK modes, it does leave an imprint on TeV scale phenomenology.
The field content of the lightest neutralino (bino versus Higgsino)
changes with $L^{-1}$ and therefore so do the couplings to matter.  
In addition, while the gaugino spectrum is approximately independent 
of scale, the scalar spectrum is not, thus this model is distinguishable
from a minimal supergravity model if $L^{-1} \ll M_{\rm Planck}$.
In fact, by measuring the scalar spectrum (e.g., at the NLC) one may
be able to determine the scale at which the scalar masses unify and 
thus the size of the extra dimensions!

\vskip.3in

{\it \underline{Note added:}} While this work was in progress we learned
that similar ideas are being pursued independently by Chacko, Luty,
Nelson, and Pont\'{o}n \cite{CLNP}.
   
\acknowledgements
Discussions with N. Arkani-Hamed, A.G. Cohen, A. de Gouv\^{e}a, J.L. Feng, 
A. Friedland, Z. Ligeti, H. Murayama, R. Sundrum, C.E.M. Wagner were helpful 
and fun.  Thanks!  We also thank the theory group at Fermilab, where this 
work was initiated, for its kind hospitality.
DEK is supported in part by the DOE under contracts
DE-FG02-90ER40560 and W-31-109-ENG-38.
GDK is supported in part by the DOE under contract
DOE-ER-40682-143.
MS is supported by the DOE under contract
DE-AC03-76SF00515.

\def\pl#1#2#3{{\it Phys. Lett. }{\bf B#1~}(19#2)~#3}
\def\zp#1#2#3{{\it Z. Phys. }{\bf C#1~}(19#2)~#3}
\def\prl#1#2#3{{\it Phys. Rev. Lett. }{\bf #1~}(19#2)~#3}
\def\rmp#1#2#3{{\it Rev. Mod. Phys. }{\bf #1~}(19#2)~#3}
\def\prep#1#2#3{{\it Phys. Rep. }{\bf #1~}(19#2)~#3}
\def\pr#1#2#3{{\it Phys. Rev. }{\bf D#1~}(19#2)~#3}
\def\np#1#2#3{{\it Nucl. Phys. }{\bf B#1~}(19#2)~#3}
\def\xxx#1{{\tt [#1]}}

\end{document}